%% file: main.tex
\documentclass[11pt,twocolumn,conference]{IEEEtran}
\usepackage{multicol}
\usepackage{caption}
\usepackage{tikz}
\usepackage{xcolor}
\usepackage{soul}
\usepackage{amsmath}
\usepackage{graphics}
\usepackage{setspace}
\usepackage{cite}
\usepackage{latexsym}
\usepackage{float}
\usepackage{epsfig}
\usepackage{multirow}
\usepackage{cite,cases,url}
\usepackage{amssymb}
\usepackage{graphicx}
\usepackage{epstopdf}
\usepackage{balance}
\usepackage{enumerate}
\usepackage{array}
\usepackage{algorithmic}
\usepackage{algorithm}
\usepackage{subfig}
\usepackage[export]{adjustbox}
\usepackage[normalem]{ulem}
\useunder{\uline}{\ul}{}
\usepackage{xcolor,colortbl}
\definecolor{LightCyan}{rgb}{0.88,1,1}
\newcolumntype{L}{>{\centering\arraybackslash}m{5cm}}
\newcolumntype{K}{>{\centering\arraybackslash}m{6cm}}
\newcolumntype{P}{>{\centering\arraybackslash}m{2.3cm}}
\newcolumntype{M}{>{\raggedright\arraybackslash}m{2cm}}
\newcolumntype{N}{>{\raggedright\arraybackslash}m{2.5cm}}

\begin{document}

\title{5G Advanced: Wireless Channel Virtualization and 
Resource Mapping 
for Real Time Spectrum Sharing
}
\author{Walaa Alqwider, Aly Sabri Abdalla
         and Vuk Marojevic
        \\
        Department of Electrical and Computer Engineering, Mississippi State University, MS, USA \\
        Emails: \{wq27, asa298, and vuk.marojevic\}@msstate.edu 
}

\maketitle


\IEEEpeerreviewmaketitle

\begin{abstract}
\label{sec:abs}
\input{./include/abs.tex}
\end{abstract}


\section{Introduction}
\label{sec:intro}

\input{./include/intro.tex}

\section{5G NR Virtual-to-Physical Resource Mapping Background}
\label{sec:specs}
\input{./include/VRMF.tex}


\section{Proposed Virtual-to-Physical Resource Mapping Framework}

\input{./include/VPRM_framework.tex}

\section{Virtual-to-Physical Mapping Technology 
}
\label{sec:extension}
\input{./include/extension.tex}

\subsection{Control Signaling and Processing}
\label{sec:VTD}

\input{./include/VRMT.tex}

\section{
Research Directions}
\label{sec:applications}
\input{./include/applications.tex}

\section{Conclusions}
\label{sec:conclusions}

\input{./include/Conclusion.tex}

\section*{Acknowledgment}
This work was supported in part by NSF grant ECCS-2030291. 

\balance

\bibliographystyle{IEEEtran}
\bibliography{swift2}
\end{document}

%% file: include/abs.tex
The coexistence between active wireless communications and passive RF spectrum use becomes an increasingly important requirement for 
coordinated spectrum access supporting critical services. The ongoing research and technological progress are focused on effective spectrum utilization including large-scale MIMO and energy efficient and low-power communications, innovative spectrum use and management, 
and resilient spectrum sharing, just to name a few. This paper introduces a new tool for real time spectrum sharing among emerging cellular networks and passive RF sensing systems used for remote sensing and radio astronomy, among others. Specifically we propose leveraging wireless channel virtualization and propose a virtual-to-physical resource mapping framework, mapping types, and control signaling that extends the current 5G New Radio (NR) specifications. Our technology introduces minimal changes to the protocol and is meant to be transparent to the end user application. We validate the proposed technology by extending a 3GPP compliant 5G NR downlink simulator and identify further research directions where 
work is needed on designing effective ways to explicitly signal the need for spectrum or spectrum use predictions.

%% file: include/intro.tex















Radio access and management 
are 
defining 5G 
wireless networks which are expected to 
support 
increasingly diverse and challenging service demands. 
Society depends heavily on  wireless technologies for commerce, transportation, health, science, and defense, with emerging technologies such as the Internet of Things (IoT) and unmanned aerial vehicles (UAVs) 
further increasing this dependence. 
Radio astronomy (RA), remote sensing (RS), and other passive RF sensing services are also indispensable in modern society. One important RS application in Earth science and climate studies is to monitor the soil moisture, which provides scientists with critical information for effective agricultural decisions, as well as for forecasting severe weather, floods, and droughts. 
These sensors measure extremely low-power natural RF signals and cannot tolerate radio frequency interference (RFI).

The growth of active wireless systems often increases RFI experienced by passive systems. Such RFI may introduce bias in the passive system's measurements or render the passive system completely useless.
While additional spectrum is being considered 
and techniques for higher spectrum efficiencies are being developed, better spectrum reuse and sharing technologies are becoming more important than ever to enable scaling 
wireless communications services while protecting passive RF systems. 
The spectrum coexistence between wireless communications 
and sensing 
is therefore an increasingly important area of research and development (R\&D). 

\input{./include/problem.tex}



\begin{figure}[t]
    \centering
    \includegraphics[width=0.8\linewidth]{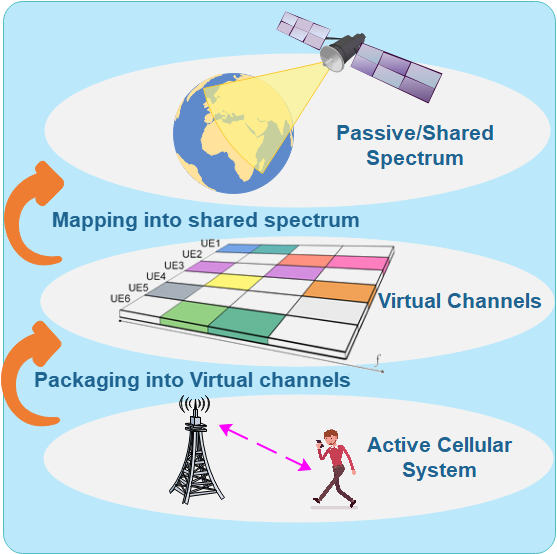}
    \caption{Spectrum coexistence enabled by wireless channel virtualization and virtual-to-physical resource mapping.}
    \label{fig:grid}
\end{figure}

This paper considers active wireless communications systems, specifically modern cellular networks, and explores a path for their evolution to enable harmonious coexistence with 
passive RF systems. 
Cellular transmissions follow a well-organized and by and large regular frame structure. This is very efficient for operation in licensed spectrum. 
We leverage the flexibility of 5G NR and 
introduce the wireless channel virtualization and a virtual-to-physical resource mapping framework 
to enable fine-grain transmission control for managing the RFI to passive RF systems. 
The proposed 
virtual-to-physical resource 
mapping 
design provides an additional layer of radio resource abstraction and 
extends the 5G NR physical layer processing chain to 
enable real-time spectrum sharing. 
Fig. \ref{fig:grid} illustrates this concept. 

The current techniques 
for sharing the spectrum among active users, mainly communications users, cannot be directly applied to
spectrum sharing between active and passive users. However, there have been recent studies that focus on enabling spectrum sharing between active and passive RF systems~\cite{MMWAVE, RFI, kaband}. 
The virtualization concept 
has became popular in wireless networking research 
to enable flexible deployment and access~\cite{5G_Spec_Virtual,SDN_Virtul_ResAllocation,LTE_Virtual_Spec_Sharing} and is a feature of the 5G core network~\cite{5GNfV,5GNFV2}. Reference~\cite{5G_Spec_Virtual} 
proposes a framework for wireless network virtualization consisting of three planes: the data plane, the cognitive plane, and the control plane. A hierarchical control scheme based on cell clustering 
is designed to coordinate the virtualized networks. 
Most of the prior works on wireless network virtualization are related to network slicing which virtualizes the network for multiple operators to access different slices of a common infrastructure/resource~\cite{floriach2018creating,slicing2020}. 

The rest of the paper is organized as follows: : Section II presents the current 3GPP specifications for supporting the virtual-to-physical mapping. Section III introduces the proposed virtual-to-physical mapping framework. Section IV shows an example use case of the proposed virtual-to-physical mapping for 5G network with different mapping types. Section V demonstrates the potential research directions that should be studied in the future work. Finally, Section VI concludes the paper with the most appealing findings.     

%% file: include/problem.tex
Radio frames of wireless protocols have a certain structure, which is more or less flexible. A 4G long-term evolution (LTE) base station is always on and transmits control channels and signals even when no users are being served. That is, the downlink frame will be sparsely populated when there are few or no users, but will use up to 25\% of the resource elements in each subframe with several control channels and signals that will appear to nearby RF sensing systems operating in the same band as strong RFI. 
5G new radio (NR) allows a more flexible configuration of channels and signals than LTE. 
A 5G base station can schedule control channels more dynamically along with user data and be put to temporary sleep%
~\cite{5GBSPower}. 
Passive RF systems, such as RA or RS, often do not have the ability to process and mitigate RFI. The problem therefore consists in modifying radio access networks for enabling coexistence with passive RF systems. 
Minimal changes to software processes as opposed to hardware redesign would enable rapid 
deployment 
as well as backward compatibility with legacy 
wireless networks.

%% file: include/VRMF.tex
\subsection{Mapping}
The virtual resources block (VRB) was introduced in 
3GPP Release 15 for 5G NR. 
It contains the modulation symbols that are mapped to a physical resource block (PRB) within the active bandwidth part (BWP) 
as part of the physical layer processing at the gNodeB. The VRB-to-PRB mapping was introduced in 5G NR to add more flexibility for radio resources allocation while decreasing the control signaling overhead. It is meant for offering frequency diversity to handle the variations in the channel quality especially for high mobility devices \cite{5gBook}.

\begin{figure*}
\centering
\vspace{-5mm}
\subfloat[5G NR non-interleaved resource mapping] 
{\includegraphics[width=3.6in]{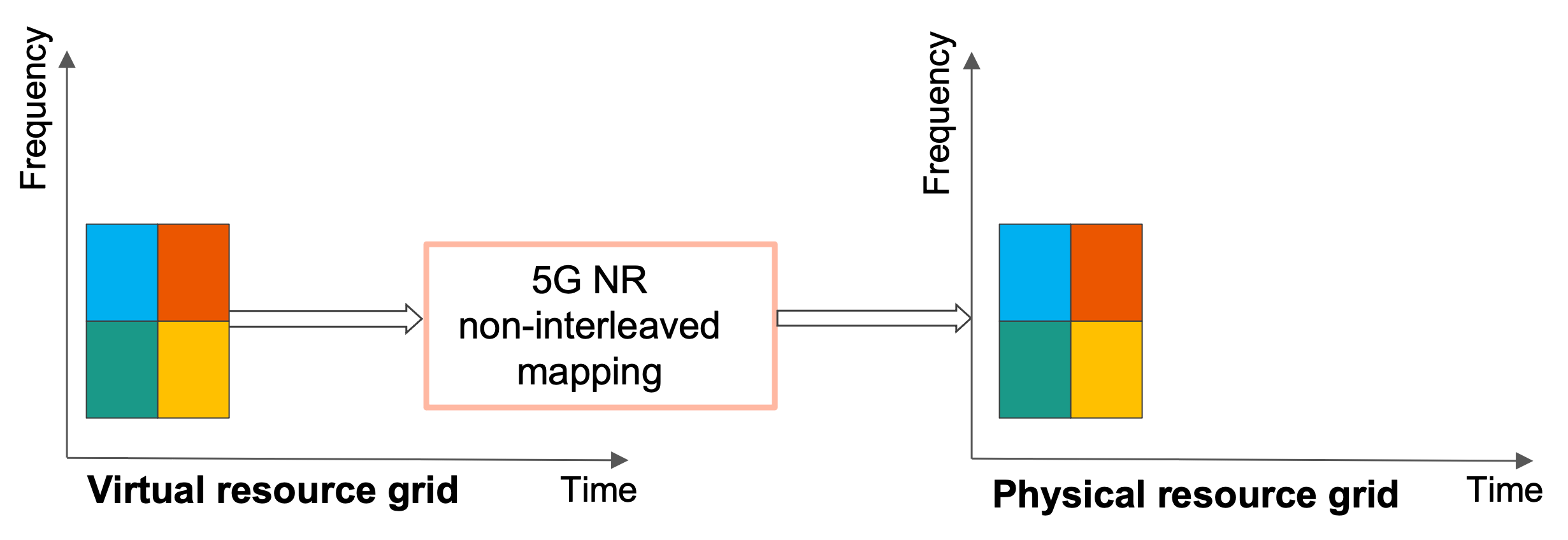}}
\subfloat[5G NR interleaved resource mapping]
{\includegraphics[width=3.6in]{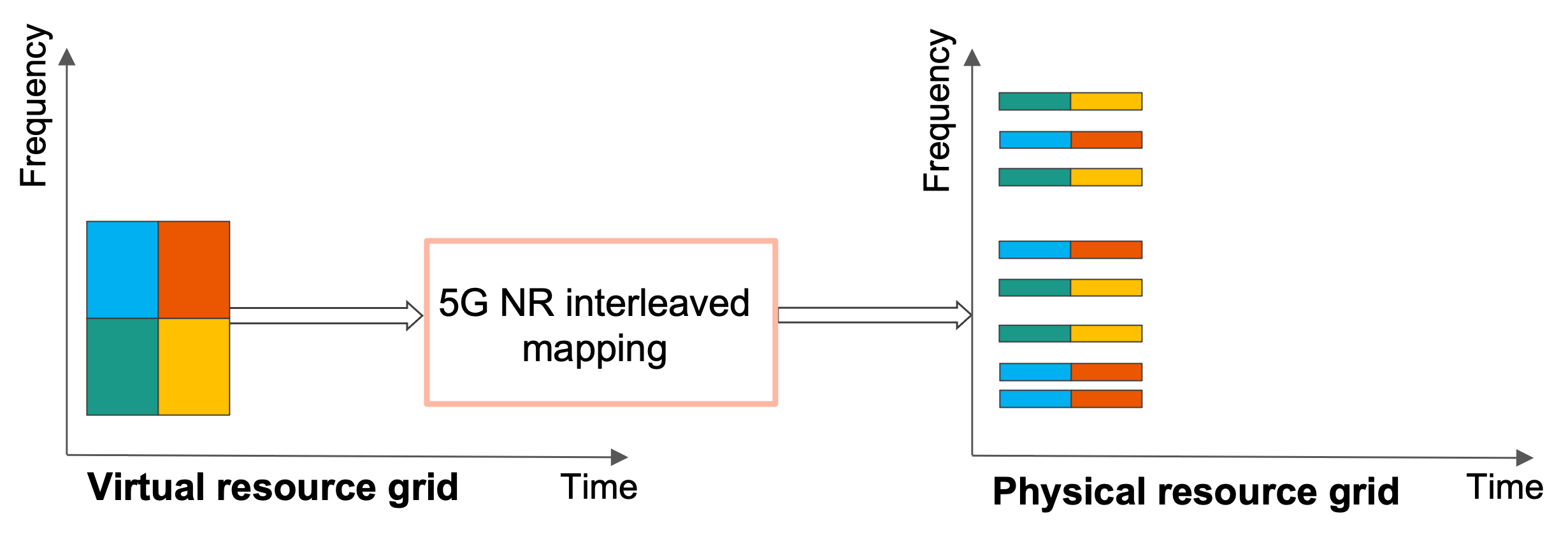}}\\
\subfloat[Proposed 5G Advanced virtual-to -physical resource mapping]
{\includegraphics[width=3.6in]{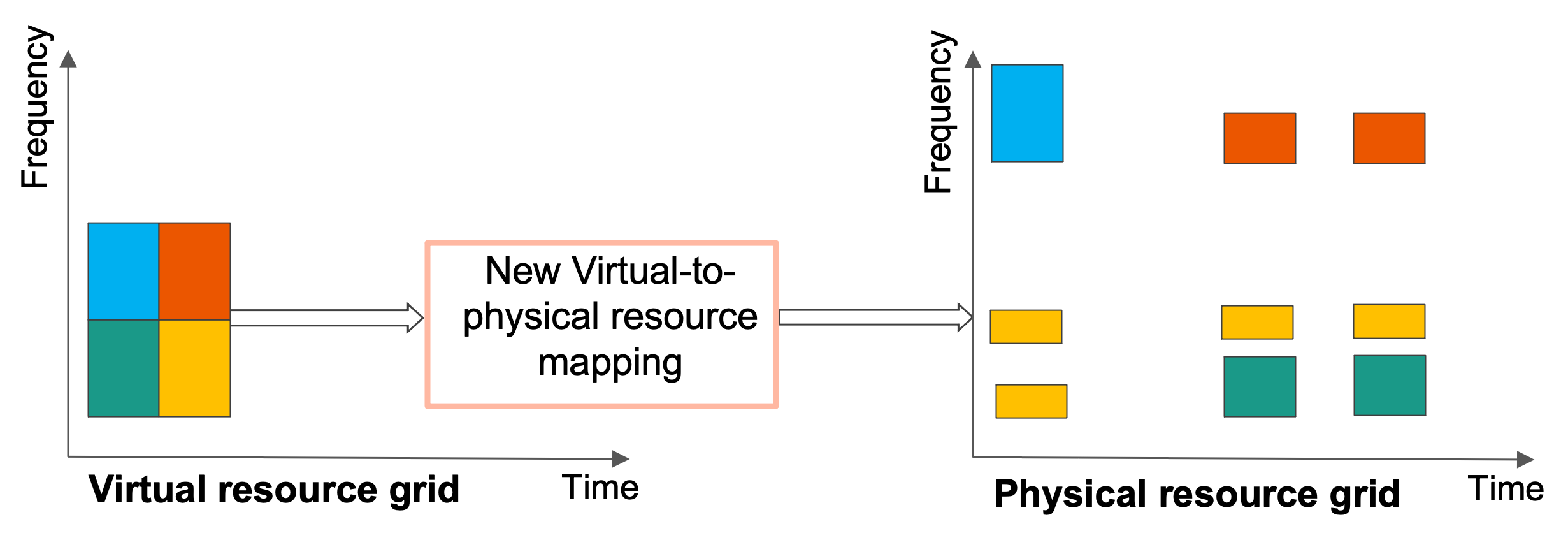}}
\caption{Virtual-to-physical resource mapping in 5G NR (a, b) and for 5G Advanced (c).}
\label{fig:mapping_typs}
\end{figure*}

The 3GPP specifications support two virtual-to-physical resource mapping types: the non-interleaved and interleaved mapping. The former maps the VRBs in the BWP directly to the same PRB in the BWP. 
The latter uses an interleaver that takes the contiguous VRBs and distributes them across the BWP. 
These two mapping options are illustrated in Figs. \ref{fig:mapping_typs}a and b. 

\subsection{Control Signaling}
The 3GPP specifications support two types of downlink resource allocation in the frequency domain. 
For both types, the gNodeB assigns VRBs to UEs, but the continuity of the allocated RBs and the number of bits used to inform the UEs about the allocation differ between the two types. Type 0 groups the available RBs in the BWP in virtual resources block groups (VRBGs). 
One or more contiguous or non-contiguous VRBGs within the BWP can be allocated to a UE. For example, 
VRBGs 1 and 5 may be allocated to a UE. 
Resources allocation Type 1, on the other hand, assigns contiguous VRBs to each UE 
\cite{data}.

The gNodeB informs the UEs about the set of RBs and orthogonal frequency division multiplexing (OFDM) symbols allocated for data transmission by encoding the scheduling information in the downlink control channel (DCI) which is carried by the Physical Downlink Control Channel (PDCCH) \cite{DCI}. Once the UE detects a valid PDCCH it extract the DCI and decodes the scheduling information to find out where to expect its downlink data 
or where it can send data on the uplink. 

The different types of DCI formats are illustrated in Table \ref{tab:DCI} \cite{DCI}. Resource allocation Type 0 employs bit mapping. That is, the number of bits 
that are needed 
to inform the UE about the allocated VRBGs is equal to the number of VRBGs within the BWP. For Type 1, the gNodeB encodes the starting virtual resource block ($RB_{start}$) and the length of the contiguously allocated RBs ($L_{RBs}$) as a resource indication value (RIV) in the resource allocation field of the DCI. This requires fewer bits than for Type 0. Type 0 thus provides more flexibility than Type 1, but it needs more control bits to inform the UEs about the RB allocation.

Non-interleaved mapping can be used for both allocation types, Type 0 and Type 1, whereas the interleaved mapping is used only with Type 1. The reason behind not using the interleaved mapping with Type 0 is that Type 0 can distribute the RBs allocated to a signle UE across the BWP for frequency diversity. Type 1 provides low-overhead 
control signaling 
where the interleaved mapping achieves the frequency diversity.  

Non-interleaved mapping is recommended with Type 1 signaling 
and a channel-dependent scheduler 
which assigns contiguous VRBs with the best channel conditions to the selected devices. Employing the interleaved mapping in this case would result 
in using RBs with worse channel conditions. 
Interleaved mapping is recommended with Type 1 in the case where scheduling decisions are not driven by the channel conditions or in high-mobility scenarios which make it difficult to follow the rapid channel variations. 
This scheme can also help with quality averaging across the transport blocks in the case where the resource allocation for one UE spans the entire BWP experiencing channel quality variations.

\begin{table}
\centering
\caption{ 5G NR DCI formats.}
\label{tab:DCI}
\begin{tabular}{|p{2cm}|p{6cm}|}
\hline
\textbf{DCI format}&\textbf{Usage}\\\hline
DCI 0\_0/0\_1/0\_2 & Schedule the Physical Uplink Shared Channel (PUSCH) in the cell \\ \hline
DCI 1\_0/1\_1/1\_2 & Schedule the Physical Downlink Shared Channel (PDSCH) in the cell \\ \hline
DCI 2\_0&Notify a group of UEs of the slot format \\ \hline
{DCI 2\_1}&{Notify a group of UEs of the PRB(s) and OFDM symbol(s) where 
no transmission is intended for them} \\ \hline
DCI 2\_2&Provide transmit power control (TPC) commands for the Physical Uplink Control Channel (PUCCH) and PUSCH\\ \hline
DCI 2\_3&Transmit a group of TPC commands for sounding reference signal (SRS) transmission by one or more UEs\\ \hline
DCI 2\_4& Notify a group of UEs of the PRB(s) and OFDM symbol(s) where to cancel the corresponding uplink transmissions 
\\ \hline
DCI 2\_5& Notify UEs about the availability of soft resources\\ \hline
DCI 2\_6& Notify one or more UEs about the power saving information outside the discontinuous reception active time
\\ \hline
DCI 3\_0 &Schedule the NR sidelink in the cell\\ \hline
DCI 3\_1 &Schedule the LTE sidelink in the cell\\ 
\hline
\end{tabular}
\end{table}

%% file: include/VPRM_framework.tex
\begin{figure}[t]
    \centering
    \includegraphics[width=0.9\linewidth]{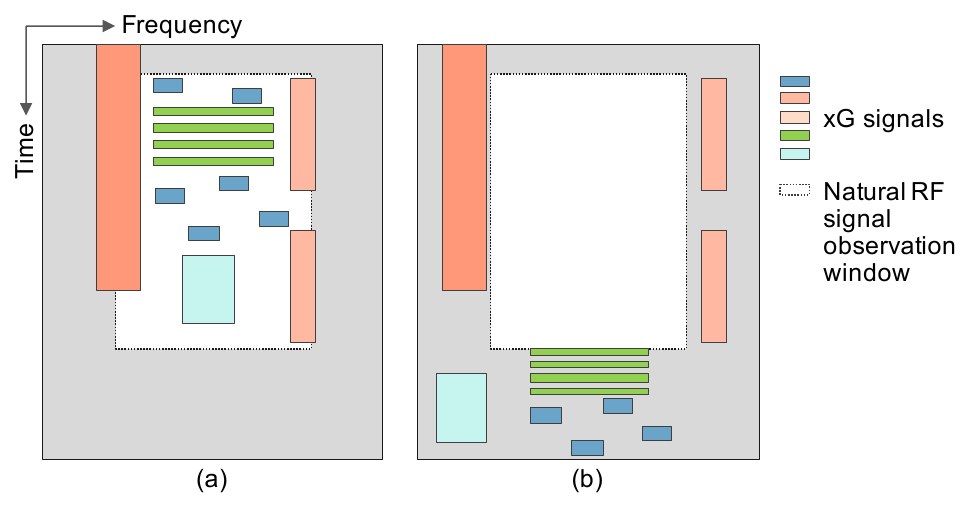}
    \caption{Active and passive RF coexistence: RFI experienced by the RF sensor(a) and RFI avoidance (b).}
    \label{fig:RFI}
\end{figure}

We propose wireless channel virtualization to control the wireless spectrum access for real-time spectrum sharing. Our virtual-to-physical resources mapping framework operates in both the frequency and time domains with the goal of avoiding active co-channel transmission during RF sensing periods. Fig. \ref{fig:RFI} illustrates this. This technology is agnostic to the wireless system and mobile network generation. 

Let us assume 
an xG network that provides regular services and continuously feeds data to its users over the wireless link. An xG transmitter packages the data and performs symbol and bit-level processing at the physical layer, using traditional digital signal processing or emerging data-driven 
techniques. Instead of directly generating the physical channels, that is, creating protocol-specific over-the-air transmission patterns, the processed transport channels become the virtual channels that use virtual spectrum and virtual radio resources instead. 
In licensed spectrum, these virtual channels may map directly to the xG physical control and data channels as per the standard 
and the scheduler. 
In shared spectrum, these channels may be 
shifted in time or frequency, or both. 
Fig.~\ref{fig:mapping_typs}c illustrates this 
type of mapping in comparison with the 5G NR mapping options of Fig.~\ref{fig:mapping_typs}a and b. In this illustrative example, a bandwidth and time segments are avoided by spreading the four collocated VRBs in frequency and time. The blue block may represent time sensitive data and is therefore transmitted without delay. 

Leveraging the fact that 
many passive radio services 
do not 
sense the same 
band continuously, we propose 
sharing this band with xG networks that implement the virtual-to-physical resources mapping for enabling flexible, on-demand transmissions, as well as silence patters to avoid harmful RFI. 
The virtual channels allocated to the xG system users will be mapped to physical channels opportunistically or systematically to minimize the 
RFI 
that the xG network may cause to other 
spectrum consumers 
and natural RF sensing systems, in particular.  
For example, the virtual channels can be grouped for uplink or downlink transmission bursts over the air, enabling silent periods, spread in both time and frequency for creating sparse OFDM grids, 
or 
partially moved to different frequencies, e.g. licensed bands, when the 
spectrum 
is needed for 
natural RF data collection. 

\begin{figure}[t]
    \centering
    \includegraphics[width=0.95\linewidth]{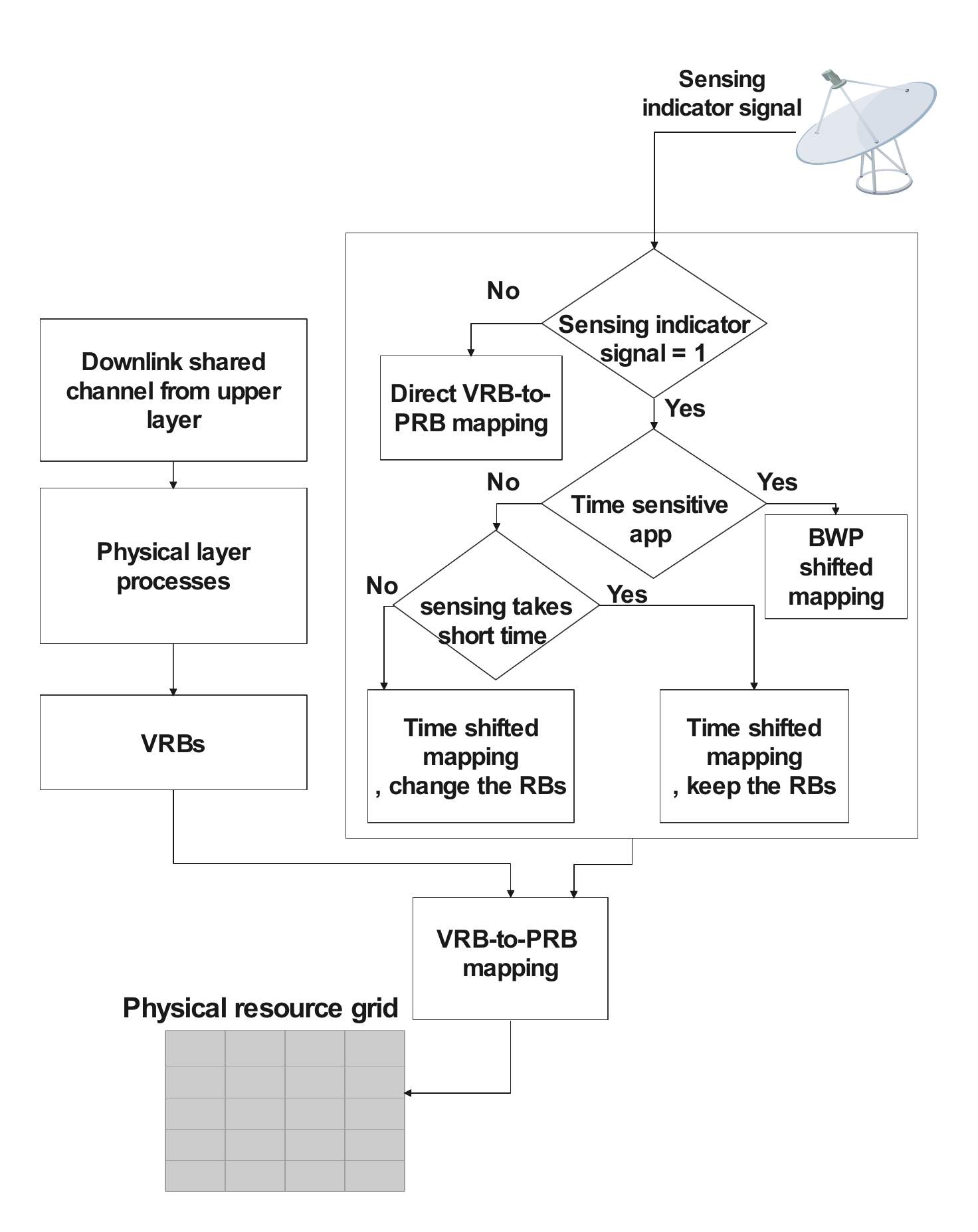}
    \caption{5G Advanced physical layer building blocks with the proposed virtual-to-physical resource mapping. 
    } 
    \label{fig:flowchart}
\end{figure}

Fig.~\ref{fig:flowchart} shows the flow of  physical layer processing in 5G NR with the proposed mapping framework. 
The base station scheduler, which is part of the medium access control 
layer, assigns VRBs to 
UEs. 
It may schedule VRBs to users in each transmission time interval (TTI) 
as a function of different inputs, 
such as channel quality indicators, 
buffer statuses, service dependent performance requirements, or other 
scheduling 
criteria. 
The data and control information 
pass through physical layer processes such as coding and modulation 
and the resource mapping combines the different streams from the physical channels according to the standard 
but maps them to the virtual resource grid, that is to virtual time-frequency resources and not the actual physical resource.
The virtual-to-physical resources mapping block is the interface or buffer between the virtual and the physical resources. It can change the physical appearance of the virtual channels and create different transmission patters. This block is followed by further processing in the digital and analog domains and the physical RF transmission. 

As indicated in Fig. \ref{fig:flowchart} the specific mapping may be determined by different factors. For example, the base station may 
receive a signal notifying about an RF sensing period in that area. 
If no sensing is expected to occur in the upcoming TTI, the direct mapping can be used. 
If there is sensing activity, the base station will consider the service that the scheduled data supports and 
if it is time sensitive and  cannot be delayed, these VRBs will be mapped 
to a band that does not affect 
the passive receiver. If the service is not time sensitive, the base station may decide to delay the transmission. 
If the sensing period is long, the channel quality will likely change and 
the virtual-to-physical resource mapping may relocate the VRBs in the physical resource grid by means of interleaving, for example.

While the proposed mechanism may 
incur processing overhead and latency, it enhances the spectrum utilization 
by leveraging bands or areas that are currently prohibited for 
communications, because they are used by highly-sensitive passive systems that cannot tolerate RFI. 
Using virtualization provides the flexibility to adapt the physical transmission for each transmitter, band, and service, and allows considering the quality of the RF hardware, RF impairments, and out-of band emissions, to dynamically control co-channel and adjacent channel interference.

%% file: include/extension.tex
This section presents the virtual-to-physical mapping technology and signaling. 
These 
are integrated into 
MATLAB's 5G Toolbox. 
We consider resource allocation Type 1 for the VRB allocation 
where the gNodeB scheduler allocates contiguous VRBs to users in each TTI for downlink transmission. 
BWPs consist of 50 RBs and three UEs are scheduled in time slots 1, 2 and 3 as illustrated in Fig.~\ref{fig:5Gmapping}a. We present four fundamental mapping types.

\subsection{Virtual-to-Physical Mapping Types}

\subsubsection{Direct Mapping} The direct mapping allocates PRBs that match the VRBs 
without any shift in the time or frequency domain. This is shown in Fig.~\ref{fig:5Gmapping}b. and used when the spectrum is available to the 5G system. The 5G users will receive their PDSCH based on the MAC layer assignments without any changes at the physical layer.  

\subsubsection{Time Shifted Mapping} 
The second type of mapping 
buffers the data for later transmission. 
The length of the buffer can be defined in time slots 
and depends on the 
passive system needs. The allocated VRBs are mapped to PRBs with a time delay while maintaining the same structure as shown in Fig.~\ref{fig:5Gmapping}c. 
This mapping uses the same spectrum as originally scheduled, but introduces a delay that certain services can tolerate but other may not. 
For latency sensitive applications other mapping types may need to be considered. 

\subsubsection{Time and Frequency Shifted Mapping} 
When 
buffering the data for transmission in later time slots as per the mapping type 2, channel-dependent scheduling schemes may become ineffective. 
Therefore, we introduce the time and frequency shifted mapping (Fig.~\ref{fig:5Gmapping}d). It allows delaying transmission and changing the frequency allocation for the UEs for frequency diversity. 
If CQI is available, these can be used for the reallocation. Alternatively, 
historical data may be available to predict the CQI for each user. Large shifts in time as determined by the mapping may trigger the rescheduling of VRBs. 

\subsubsection{Shift to another {BWP}} 
\begin{figure*}
\centering
\vspace{-5mm}
\subfloat[VRB allocation] 
{\includegraphics[width=2.44in]{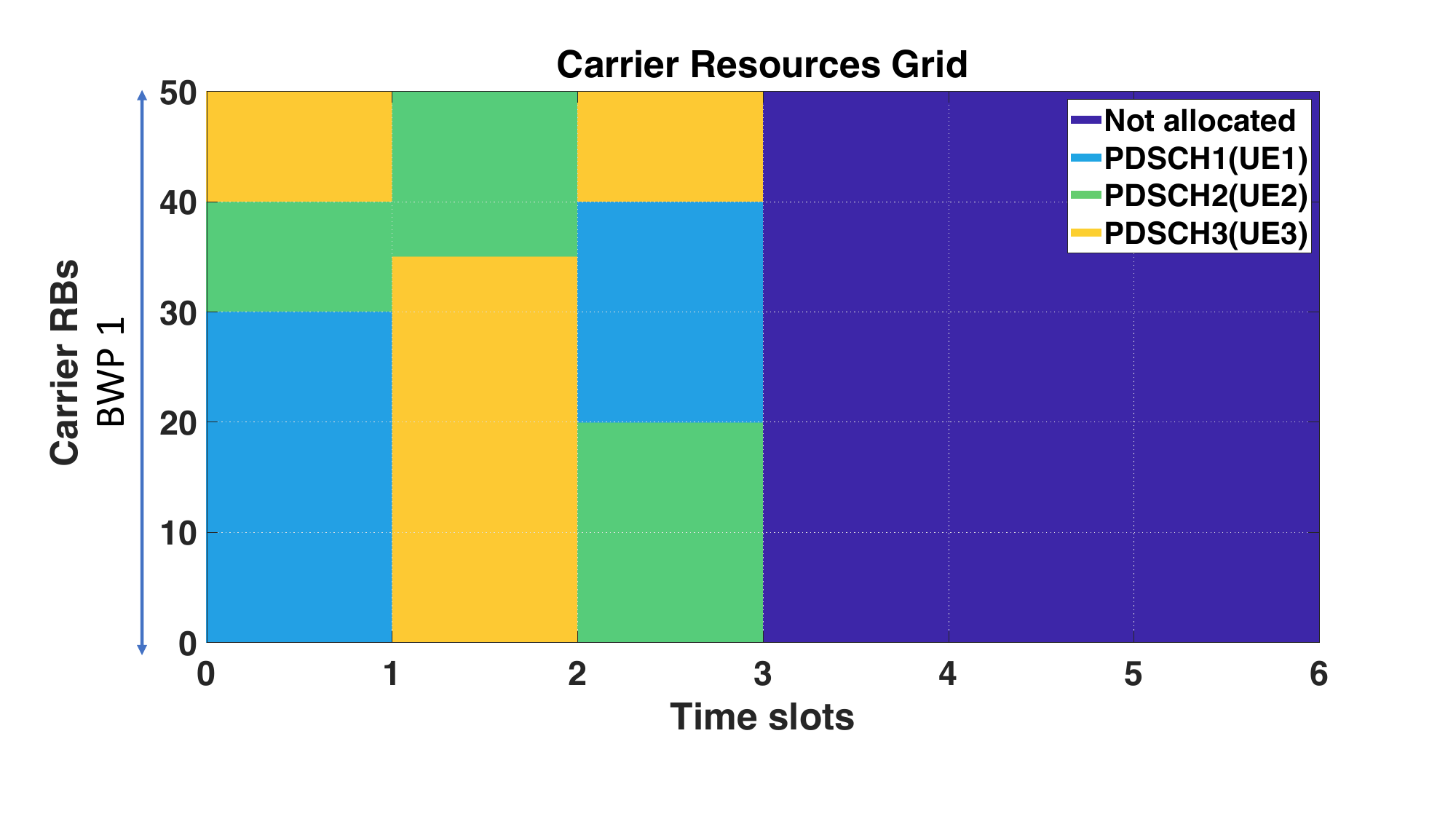}}
\subfloat[Type 1: Direct mapping]
{\includegraphics[width=2.44in]{figures/Examples_5G/original_5G.pdf}}
\subfloat[Type 2: Time shifted mapping]
{\includegraphics[width=2.44in]{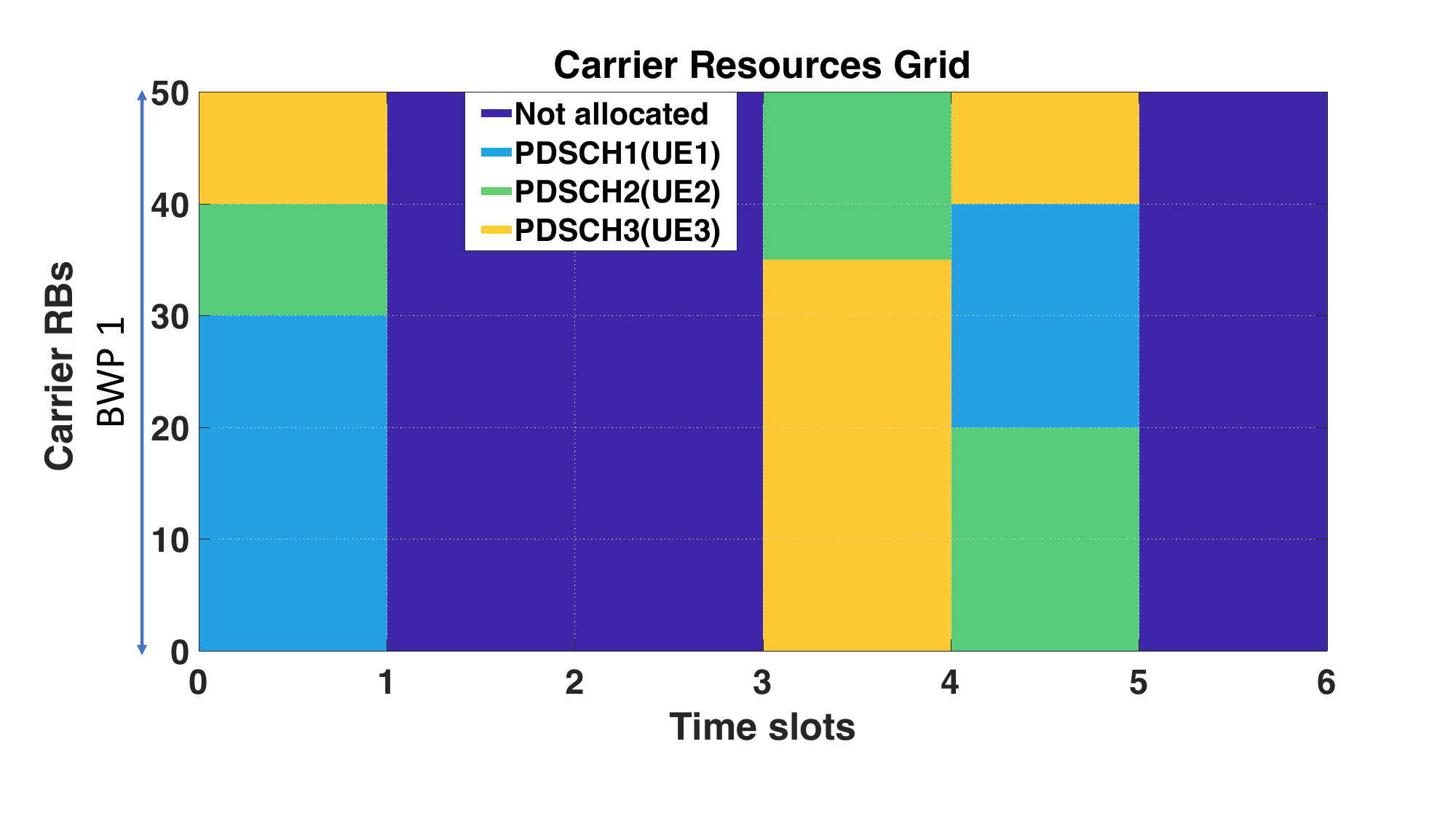}}\\
\subfloat[Type 3: Time and frequency shifted mapping ]
{\includegraphics[width=2.44in]{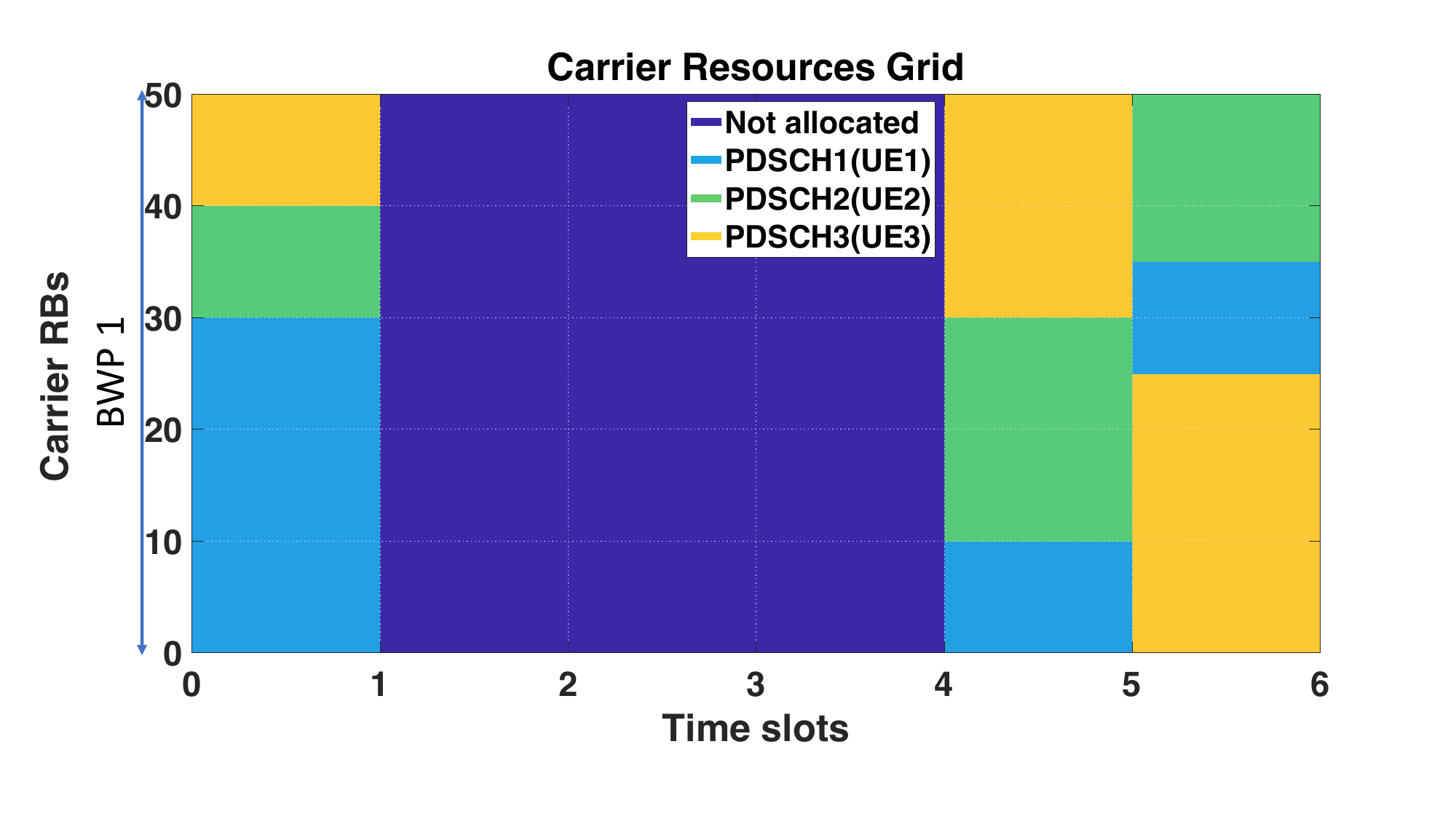}}
\subfloat[Type 4: BWP shifted mapping]
{\includegraphics[width=2.44in]{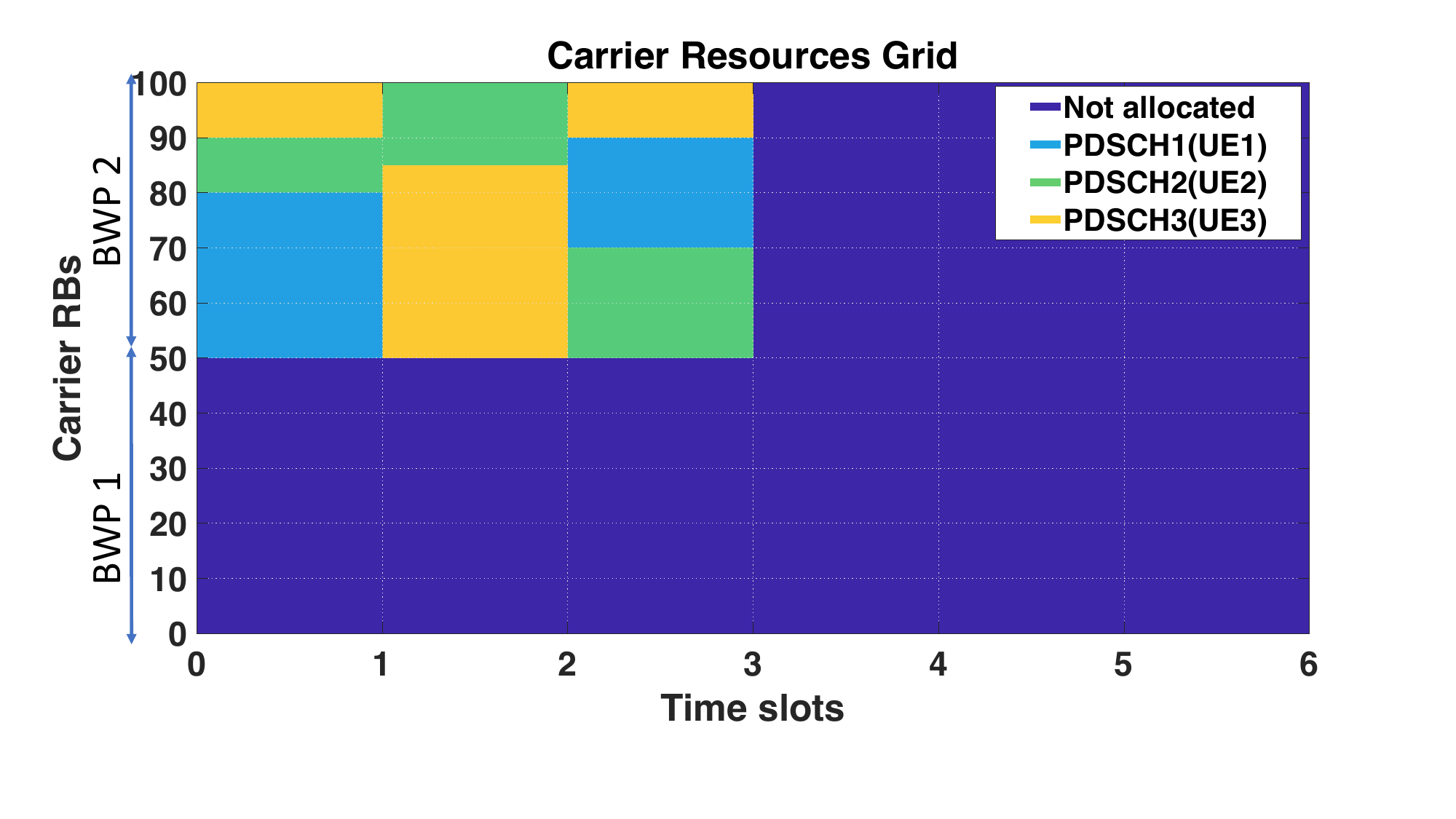}}
\caption{Examples of different types of mapping for 5G NR waveform}
\label{fig:5Gmapping}
\end{figure*}

When delaying the transmission is not suitable, for example, for time sensitive data, a frequency shifted mapping is proposed and is illustrated in Fig.~\ref{fig:5Gmapping}e. 
We assume the case of the 5G NR network, the allocated frequency is divided in BWPs. 
Therefore, to free the current BWP for the passive device, the mapping block shifts the allocated VRBs to another available BWP and informs the users about this frequency shift by sending a signal to activate the new BWP. The advantage of this mapping is that no latency is incurred which is needed when latency is the critical metric. The new BWP may within the carrier licensed spectrum and the original BWP may be in the shared spectrum that cannot be momentarily used for communications. 


%% file: include/VRMT.tex
The proposed virtual-to-physical resource mapping, or VRB-to-PRB mapping, needs special control signals to inform the active users about the mapping so that these users can look for there downlink data in the correct place in the frequency and time domains. This applies for the uplink as well, but it may be the base station that controls it. For the sake of brevity, we discuss the downlink here. 

The base station needs to send a control signal to inform the user which mapping type it uses. As we have 4 different types we can use a control signal with two bits (00, 01, 10, 11) to indicate 
direct mapping, time shifted mapping, time and frequency shifted mapping, and BWP shifted mapping, respectively. 
No additional more control signal is needed for the first mapping type because the active users will receive the downlink data in the same frequency and time resources as signaled in the PDCCH informing the users about their scheduled resources. 
For the mapping type 2, the active users need to know the delay of the downlink data. The base station thus needs to signal the time shift. 
For mapping type 3, which is the most complex type, the new frequency and time allocations need to be signaled. For mapping type 4, if an active user uses more than 2 BWPs it needs to know in which BWP it should expect the data and, therefore, additional control signaling is needed for this type to indicate the BWP. 
Such signaling needs to happen as soon as the mapping is determined to inform the receivers for coherent demodulation and for potential power saving. 

Section \ref{sec:specs} introduced the DCI formats that are used in 5G NR to carry downlink scheduling assignments, uplink scheduling grants, and other control information for one user or a group of users. The scheduling information is used to inform a UE which RBs in the frequency domain and where in the time domain 
to expect downlink data on the PDSCH or where to send uplink data on 
the PUSCH. The DCI data is attached with a 24-bit cyclic redundancy check (CRC) which is scrambled with the corresponding identifier referred to as a radio network temporary identifier (RNTI). Using the RNTI mask, the UE can detect the DCI for its unicast data and distinguish sets of DCIs with different purposes that have the same payload size. 
The UE might receive the DCI information in the same slot where it intends to receive the data. This is called self-contained slot because it contains both the scheduling information and the data.

We propose leveraging the existing 
DCI formats to signal 
about the changes 
in the physical resources allocation. For the first mapping type, no additional signaling is needed since the UE received the data as determined by the scheduler and already encapsulated in the DCI as per the 3GPP standards. 
For all other mapping types introduced above we can make use of the self-contained slot principle. 
The gNodeB can use DCI 2\_1 to notify the group of UEs that (a) the allocated resources are transmitted on another 
BWP in case of mapping type 4, or (b) 
indicate the number of slots that the UEs will not receive any data on for mapping types 2 and 3. 
Informing the UEs about the empty slots where no data is transmitted 
will help improve the energy consumption as the UEs can go to idle state during these time slots.
The DCI 2\_1 format can be modified to inform the UEs about the type of mapping used and any other essential information associated with this mapping type, such as the new BWP in case of mapping type 4, the number of shifted time slots in case of mapping types 2 and 3. For mapping type 3 where the PRBs will not match the VRBs shifted in time, 
the gNodeB 
can encode the changes in PRBs 
using DCI 1\_0/1\_1 for the case of downlink transmission. 

%% file: include/applications.tex
The proposed wireless channel virtualization and virtual-to-physical resource mapping provides a new tool for real time spectrum sharing. More research, prototyping, and testing is needed. We recommend the following research directions. 

\begin{itemize}
    \item \textbf{Passive system feedback mechanisms:} Generally there are different options for informing about sensing periods. The data can be available beforehand and stored in a database that is accessible to the active transmission system. Oftentimes, this is not feasible because of privacy and security concerns. For real time spectrum sharing, we propose other types of control signaling for coordinating active and passive use of spectrum. 
    Different types of sensors may offer different solutions as a function of location and sensing patterns. RF sensors such as the SMAP satellites currently provide global RFI information where RFI was detected in the past. 
    Research needs to study active feedback mechanisms or predictions before RFI happens.  
    \item \textbf{Coordinated virtual-to-physical resources mapping:} We suggest research on data driven learning approaches 
    to create more dynamic, efficient, and coordinated mapping between the virtual to physical resources. There are a limited number of major passive sensing systems at static and well-known locations and their 
    frequencies and sensing patterns may be known or 
    predicted. It may also be possible to evaluate RFI from collected and processed data at the sensing sites to create 
    and maintain interference maps. Such maps can be dynamically updated and help finding the optimal mapping opportunities between the virtual and physical resources. When done 
    proactively, e.g. through tighter collaboration with passive RF systems and leveraging advanced computing and communications principles, such as caching or prefetching of data, 
    real-time performance can be improved for active communications while minimizing RFI to passive users.   
    
    
    
    \item \textbf{Cross-layer coordination:} The link and system performance when applying virtual-to-physical resource mapping needs to be evaluated in different radio environments. Specifically, for frequency selective channels and where the time shift is larger than the channel coherence time, schedulers may need to be redesigned or scheduling decisions reevaluated. Similarly, precoding for MIMO and beamforming 
    may be suboptimal if CSI is outdated. These and other channel-driven processes therefore have to be evaluated for the introduced frequency and time agility of the proposed transceiver.

    \item \textbf{
    Frequency-selective 
    waveform design:} It is possible to design waveforms that embed certain features, e.g. cyclostationary signal features, that allow message decoding even after removing parts of the signal before transmission that would otherwise cause strong RFI. 
    This approach is challenging as it needs careful consideration of transceiver characteristics, including processing complexity 
    and peak-to-average Power ratio 
    characteristics. 
    The design of such waveforms need to consider the characteristics of the passive system, its observations, and the equipment used. 
    This can be explored an an alternative technology to the one proposed in this paper or achieved through a custom virtual-to-physical mapping block.   
    
    
    \item \textbf{Machine learning-enhanced mapping:} Machine learning (ML) and, specifically, reinforcement learning (RL) should be explored for making a cleaver mapping decisions based on predictions. For example, 
    ML can predict the channel quality for the different users in different PRBs for the upcoming time slots by providing 
    a series of the  past CQIs for the different users to estimate the expected 
    CQIs for the next time slots. The output of the ML CQI predictor will be fed to the RL algorithm in order in to decide which VRBs to mapped to which PRBs to optimize the use of resources for communication when they are available. This method can run while the passive receiver is using the spectrum. 
\end{itemize}

%% file: include/Conclusion.tex
This paper has introduced wireless channel virtualization and a virtual-to-physical resource mapping framework for the spectrum coexistence between active and passive RF systems. Specifically, we consider 5G and extend its current protocol features to enable real time spectrum sharing. We introduce a layer of abstraction at the physical layer to dynamically control access to physical radio resource in time and frequency. 
We introduce several mapping types and control signaling that leverage and extend the current 5G NR specifications. Our technology introduces minimal changes to the protocol and is meant to be transparent to the end user application. We validate the proposed technology by extending a 3GPP compliant 5G NR downlink simulator and identify further research directions where 
work is needed on designing effective ways to explicitly signal the need for spectrum or spectrum use predictions.